\input{aipcheck}

\documentclass[final]{aipproc}

\layoutstyle{8x11double}

\newcommand{\abs}[1]{\left\vert#1\right\vert}

\newcommand{\ket}[1]{\vert #1 \rangle}

\begin{document}

\title{Solid State Implementation of Quantum Random Walks on General Graphs}



\author{K. Manouchehri} {}

\author{J. B. Wang} {
  address={School of Physics, The University of Western Australia, Perth 6009,
  Australia},
  email={wang@physics.uwa.edu.au}
}

\begin{abstract}
    Advances in recent years have made it possible to explore quantum dots
    as a viable technology for scalable quantum information processing.
    Charge qubits for example can be realized in the lowest bound states of
    coupled quantum dots and the precision control of the confinement
    potential allows for the realization of a full set of universal
    qubit gates, including arbitrary single-qubit rotations and
    two-qubit C-NOT gates. In this work we describe a novel scheme for
    implementing quantum random walks on arbitrarily complex graphs
    by extending these elementary operations to the control of a two-dimensional
    quantum dot grid. As single-qubit rotations constitute the
    essential building blocks of our implementation scheme, we also present
    numerical simulations of one such mechanism by directly solving the
    corresponding time-dependent Schr\"{o}dinger equation.
\end{abstract}

\maketitle

\section{Introduction}

The remarkable speed-up observed in the Shor factorization algorithm
\cite{Shor1997} and the Grover search algorithm \cite{Grover1997} has
prompt intense interest in developing appropriate physical systems
for quantum computation and quantum information processing. Among the
various proposals, solid-state systems are particularly attractive,
since they are potentially ready to be integrated into large quantum
networks and are enticing to the present semiconductor industry.

\emph{Quantum Dots} \cite{DiVincenzo2005} have in particular inspired
many solid-state based proposals for quantum information processing
and are routinely manufactured in numerous laboratories. When
confining exactly one electron, pairs of coupled quantum dots can
naturally form qubits, the building blocks for quantum circuits.
Charge qubits for example can be formally defined as $\alpha\ket{L}
+ \beta\ket{R}$ where $\alpha$ and $\beta$ represent the single
electron amplitude to be present in the the lowest bound states of
the left and right quantum dot respectively. There have been a number
of proposals for performing quantum operations on solid state charge
qubits \cite{Fedichkin2000, Tanamoto2000, Bertoni2000, Hollenberg2004}. In particular references
\cite{Green2004, Dinneen2008} have demonstrated that a precision
control of the confinement barrier potential allows for the
realization of arbitrary single-qubit rotations.

In this paper we utilize these elementary qubit operations as the
basis of a novel scheme for physically implementing quantum random
walks on any arbitrary undirected graph. To do so we will first
provide a theoretical platform for quantum random walks on graphs
followed by a description of our proposed physical implementation
scheme. Finally we provide the results of our numerical simulations
for performing robust qubit rotations, which are fundamental for
implementing of the walk.

\section{Quantum Walk on a Graph} \label{section.qrw-theory}

\subsection{Motivation}

Random walks have been employed in virtually every science related
discipline to model everyday phenomena such as the DNA synapsis
\cite{Sessionsa1997}, animals' foraging strategies
\cite{Benichou2005}, diffusion and mobility in materials
\cite{Trautt2006} and exchange rate forecast \cite{Kilian2003}. They
have also found algorithmic applications, for example, in solving
differential equations \cite{Hoshino1971}, quantum monte carlo for
solving the many body Schr\"{o}dinger equation \cite{Ceperley1986},
optimization \cite{Berg1993}, clustering and classification
\cite{Scholl2003}, fractal theory \cite{Anteneodo2007} or even
estimating the relative sizes of Google, MSN and Yahoo search engines
\cite{Bar-Yossef2006}.

Whilst the so called \emph{classical} random walks have been
successfully utilized in such a diverse range of applications,
\emph{quantum} random walks are expected to provide us with a new
paradigm for solving many practical problems more efficiently
\cite{Aharonov1993, Knight2003-1}. In fact quantum walks have already
inspired efficient algorithms with applications in connectivity and
graph theory \cite{Kempe2003, Douglas2008}, as well as quantum search
and element distinctness \cite{Shenvi2003, Childs2004}, due to their
non-intuitive and markedly different properties, including faster
\emph{mixing} and \emph{hitting} times.

\begin{figure}[htb]
    \includegraphics[width=7cm, bb=0 0 520 810,clip]{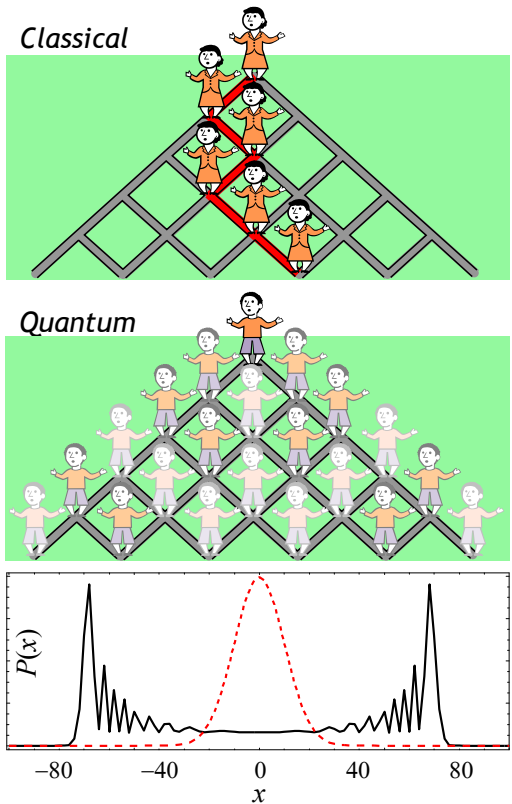}
    \caption{A comparison of quantum vs. classical random walk.
    A classical random walk leads to a Gaussian distribution (dotted) while
    its quantum counterpart produces an un-intuitive distribution (solid) which
    expands very rapidly.}
    \label{fig.qrw-intro}
\end{figure}

To illustrate, take Alice, who is a classical random walker
originally positioned at $x=0$. To decide whether to take a step to
the left or right, he flips a coin with the two possible outcomes
labeled by $+$ and $-$ and respective probabilities $P_+$ and $P_-$.
Looking at the outcome, she then knows with certainty which way to
move; either to the left ($x=-1$) \emph{or} to the right ($x=+1$).
Hence after many iterations of this process Alice's classical walk
traces a single path within a decision tree and the probability for
finding him at a given position $x$ follows a Gaussian distribution.
Meanwhile Bob, who is a quantum walker, flips his coin but never
looks at the outcome. Instead he steps simultaneously to the left
\emph{and} right with complex amplitudes $\mathcal{A}_+$ and
$\mathcal{A}_-$ such that $\abs{\mathcal{A}_+}^2=P_+$ and
$\abs{\mathcal{A}_-}^2=P_-$. After many interactions Bob's quantum
walk results in a probability wavefunction with a finite amplitude to
be present everywhere in the tree. At the end we can get Bob back
in one piece by ``collapsing'' his wavefunction. This will allow him
to emerge at some position $x$ with a probability given by the
peculiar distribution depicted in Fig \ref{fig.qrw-intro}.

Over the last few years there have been several proposals for such a
physical implementation using a variety of physical systems including
NMR \cite{Ryan2004}, cavity QED \cite{Di2004,
Agarwal2005}, ion traps \cite{Travaglione2002}, classical and quantum
optics \cite{Knight2003-1, Francisco2006, Zou2006, Zhang2007,
Kovsik2005}, optical lattice and microtraps \cite{Joo2006,
Eckert2005} as well as quantum dots \cite{Manouchehri2008,
Solenov2006}. None of the existing proposals however consider quantum
random walks on general graphs, with the majority describing only a
one-dimensional implementation. This is while from an application
point of view most useful algorithms would involve traversing graphs
with arbitrarily complex structures. In this paper, we build on an
earlier implementation scheme based on optical lattices
\cite{Manouchehri2008-2}, and advance a proposal for realizing a
universal quantum random walk capable of traversing any general
undirected graph using a two-dimensional grid of quantum dots.

\subsection{Theory}

Let us first consider a complete graph with all possible
connections between the $\mathcal{N}$ nodes including self loops. We
will relax this constraint later by removing the unwanted
connections. Here the walker requires an $\mathcal{N}$-sided coin for
moving from one node to $\mathcal{N}$ other nodes. The complete state
of the walker is therefore described by $\ket{\psi} =
\sum_{j=1}^\mathcal{N} \sum_{k=1}^\mathcal{N}
\mathcal{A}_{j,k}\ket{j,k}$, where $\ket{j}$ denotes the nodes or
position states of the walker, $\ket{k}$ specifies the state of the
coin, and $\mathcal{A}_{j,k}$ is a complex amplitude. A coin flip in
the context of this quantum walk corresponds to a unitary rotation of
the coin states at every node $j$ using an
$\mathcal{N}\times\mathcal{N}$ matrix $\hat{c}_j$ also known as the
coin operator. The coin operation is followed by the walker stepping
from node $j$ simultaneously to all other nodes on the graph using a
conditional translation operator $\hat{T}$ such that
$\hat{T}\ket{j,k}\longrightarrow\ket{j',k'}$ according to some
predefined rule, where $j$ and $j'$ label the two nodes at the end of
an edge $e_{jj'}$ \cite{Kendon2005}. The quantum walk evolves via
repeated applications of the coin followed by the translation
operator. More explicitly
\begin{equation} \label{eqn.qrw-evolution}
    \ket{\psi_n} = \hat{T}_n~\hat{\mathcal{C}}_n \ldots\hat{T}_2~\hat{\mathcal{C}}_2~\hat{T}_1~\hat{\mathcal{C}}_1 ~
    \ket{\psi_0},
\end{equation}
where $\ket{\psi_0}$ is the initial state of the walker,
$\ket{\psi_n}$ corresponds to the state of the walk after $n$ steps,
$\mathcal{N}^2\times\mathcal{N}^2$ matrices $\hat{\mathcal{C}}_i$ and
$\hat{T}_i$ are the coin and translation operators at the $i$th step,
and $\hat{\mathcal{C}}$ incorporates the individual coin operators
$\hat{c}_1 \ldots \hat{c}_\mathcal{N}$ which simultaneously act on
all the nodes. The operators $\hat{c}$ can in principle invoke
different rotations at each node $j$, but are often uniformly set to
be the Hadamard matrix.

\begin{figure}[htb]
    \includegraphics[width=7cm, bb=0 0 430 700,clip]{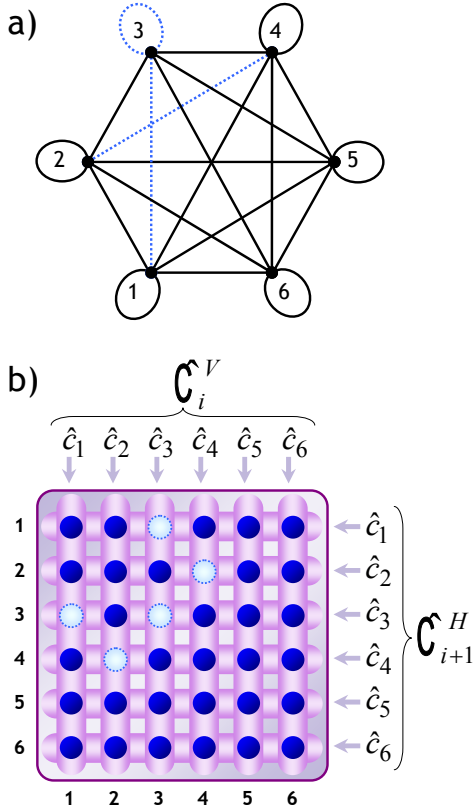}
    \caption{a) A complete graph with $\mathcal{N}=6$ nodes. b) The quantum walk
    Hilbert space viewed as a 2D array, where the element $(j, k)$ represents the state $k$
    under node $j$. A generalized graph can be constructed by removing edges (dotted
	lines) from a complete graph. This can be physically achieved by modifying the coin operators in
	such a way as to isolate the unwanted states (dotted circles) from interacting with other states.}
    \label{fig.qrw-on-graph}
\end{figure}

In \cite{Manouchehri2008-2} we have shown that for any arbitrary
graph the Hilbert space of the walk can in fact be represented as a
two-dimensional $\mathcal{N} \times\mathcal{N}$ grid (Fig
\ref{fig.qrw-on-graph}) where each individual node on the graph
corresponds to a grid row (column) and the coin states within that
node are the individual grid elements along that row (column). There,
it is also shown that the quantum walk
evolution given in Eq. \ref{eqn.qrw-evolution} can effectively be
reduced to
\begin{equation} \label{eqn.qrw-evolution-2}
    \ket{\psi_n} = \hat{\mathcal{C}}_n^V \hat{\mathcal{C}}_{n-1}^H \ldots \hat{\mathcal{C}}_2^V~\hat{\mathcal{C}}_1^H ~
    \ket{\psi_0},
\end{equation}
where $\hat{\mathcal{C}}_i^H$ ($\hat{\mathcal{C}}_i^V$) correspond to
the application of the coin operator for the $i$th step, on the grid elements that are grouped horizontally (vertically) as
depicted in Fig. \ref{fig.qrw-on-graph}.

Any general graph can now be constructed from its corresponding
complete graph by removing the unwanted edges as depicted in Fig.
\ref{fig.qrw-on-graph}. Eliminating an edge $e_{jj'}$ corresponds to
eliminating the connection between two states $\ket{j,j'}$ and
$\ket{j',j}$. However instead of removing these states from the
Hilbert space, the action of the coin operators $\hat{c}_i^H$
($\hat{c}_i^V$) can be altered in such a way as to isolate the
unwanted states from interacting with other states.

\section{Physical Implementation} \label{section.qrw-implementation}

We propose utilizing a two-dimensional grid of quantum dots to
represent the quantum walk's Hilbert space, where the initial
distribution of a single electron wavefuncion among the dots
represents the initial state of the walk. As we will see this scheme
necessitates the construction of a $2\mathcal{N} \times 2\mathcal{N}$
grid, where every second row (column) of dots will be used as
temporary ``register''. In what follows we present a mechanism for
manipulating the electronic wave function throughout the grid by
introducing appropriate quantum dot interactions in a manner which
corresponds exactly to the quantum random walk evolution described in
Eg. \ref{eqn.qrw-evolution-2}. Naturally the resulting electron
probability wave distribution gives the final state of the quantum
walk.

Considering the evolution of the walk in Eq.
\ref{eqn.qrw-evolution-2}, what we require is a means by which we can
implement the action an arbitrary $\mathcal{N}$-level unitary
operator $\hat{c}_j^H$ ($\hat{c}_j^V$) on the $\mathcal{N}$ nodes
along the $j$th row (column) of the grid. In \cite{Manouchehri2008-2}
we showed that this can be achieved via a CS decomposition
\cite{Edelman2007} which essentially reduces a general
$\mathcal{N}$-level rotation matrix to a series of pair-wise or qubit
rotations which can in principle be readily implemented. What makes
this implementation scheme non-trivial however is the fact that the
resulting pair-wise interactions are not limited to neighboring
nodes. More precisely it can be shown that for a general rotation of
column $j$ for example, we have
\begin{equation}\label{eqn.cs-decomposition}
    \hat{c}_j^v = \prod_{i = 1}^{\mathcal{N} - 1} \mathcal{U}_i(d),
\end{equation}
for $d \in [2, 4, \ldots \mathcal{N}/2]$, where the action of each
$\mathcal{U}_i(d)$ on the nodes of row $j$ consists of
$\mathcal{N}/2$ simultaneous pair-wise interactions between nodes $k
d+r$ and $k d+r+d/2$ where $k=0,\ldots,\mathcal{N}/d-1$, $d=2,
4,\ldots \mathcal{N}/2$ and $r=1, 2 \cdots d/2$. Clearly for all
$d\neq2$, interactions are non-neighboring but follow a systematic
form.

In the following sections we describe two mechanisms: 1) How to
implement a pair-wise interaction between any two neighboring quantum
dots, and 2) How to extend this to non-neighboring quantum dots.

\subsection{Pair-wise Interactions}

\begin{figure}[htb]
    \includegraphics[width=7cm, bb=0 25 420 240,clip]{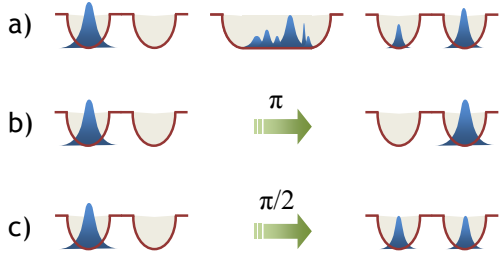}
    \caption{a) Arbitrary qubit rotations can be performed by a precision control
    of the central potential barrier. b) The barrier raised in time to implement a $\pi$
    rotation. c) the barrier is raised in time to implement a $\pi/2$ rotation.}
    \label{fig.qubit-rotation-scheme}
\end{figure}

A pair of neighboring quantum dots can be made to undergo a unitary
rotation $\hat{R}$ by the precision control of the potential barrier
between them \cite{Green2004, Dinneen2008}. Assuming that the
confined electron is initially in the left dot for example (Fig.
\ref{fig.qubit-rotation-scheme}), by dropping the barrier the
electron wave packet is free to move between the two dots. Returning
the barrier to its initial state at a precise moment in time makes it
possible to recapture the electron but this time with the desired
distribution across the two dots. We have presented a numerical
simulation of this scheme in the last section.

Here we adopt the usual terminology to describe a $\pi$ rotation as
one in which the electron wave-packet is transferred entirely from
one dot to its neighboring dot. This is also equivalent to a SWAP or
NOT gate. Likewise a $\pi/2$ rotation represents a $50-50$ split of
the wave-packet initially confined to either of the dots.

\subsection{Quantum Dot Conveyor Belt}

To interact a pair of non-neighboring quantum dots we are effectively
forced to move the two dots close to each other, apply the desired
rotation and then return them to their original location.

\begin{figure}[htb]
    \includegraphics[width=7cm, bb=0 0 410 740,clip]{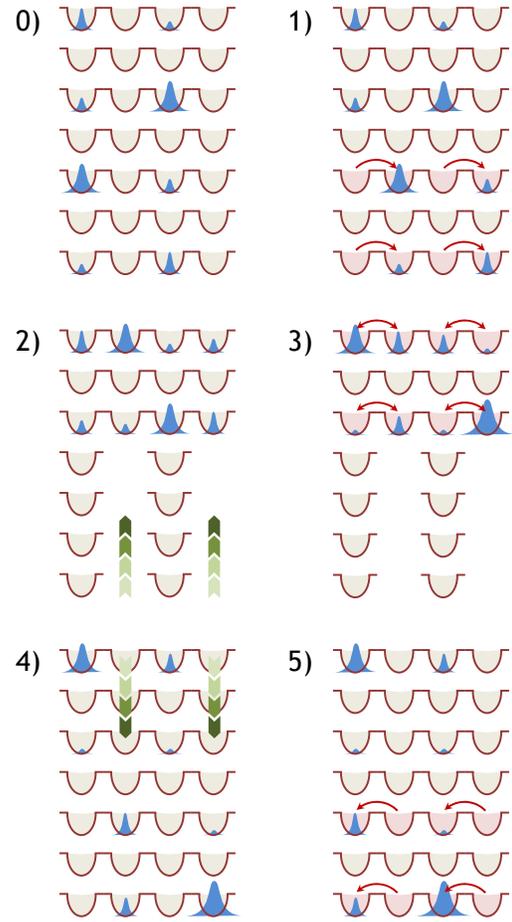}
    \caption{The initial electron wavefunction inside the quantum dot grid
    followed by 5 steps required to implement the operator $\mathcal{U}_i(d=4)$.}
    \label{fig.qrw-operation}
\end{figure}

What makes this process systematic and practically viable is the
specific pattern of pairwise interactions arising from Eq.
\ref{eqn.cs-decomposition}. As depicted in Fig.
\ref{fig.qrw-operation}, to implement the action of each
$\mathcal{U}_i(d)$ we take the following five steps:
\begin{enumerate}
  \item Apply $\mathcal{N}/2$ simultaneous $\pi$
rotations to pairs of quantum dots at $k d+r$ and their adjacent
register dot. This has the effect of transferring the electron wave
packets to the register dots.
  \item Adiabatically move the register
quantum dots, much like a "conveyor belt" carrying the electron
wavepackets along the register row (column). This can be archived by
carefully designing and manipulating the voltage applied to the
electrodes such that the confinement potentials experience an
effective motion. Moving the register dots by an amount equal to $d$
times the quantum dot width would effectively allow the amplitudes at
$k d + r$ to be coupled with the amplitudes at $k d + r + d/2$.
  \item Simultaneously apply $\mathcal{N}/2$ general rotations $\hat{R}$ to
the newly coupled quantum dot pairs.
  \item Return the register qubits to
their original location by reversing the adiabatic motion;
  \item Introduce another $\pi$ rotation to move the amplitudes from register
dots back to their original positions.
\end{enumerate}

We emphasis that the above steps can be carried out simultaneously
throughout the entire grid and the rotation $\hat{R}$ in step 3 can
be different for each quantum dot pair.

\subsection{Modeling Qubit Rotations} \label{section.qdot-simulation}

One of the key issues in quantum computation is to design a
time-dependent Hamiltonian as precisely as possible, so that one can
drive the system undergo the required logic gate manipulation. Since
the intrinsic parallelism of a quantum computer comes from
superposition and entanglement where phases and amplitudes play an
eminently important role, such a design can only be achieved through
detailed theoretical calculations including all perceptible
interactions in the system.

The Schr\"{o}dinger equation governing the quantum dynamical evolution of
few-electron systems reads
\begin{eqnarray*}
\lefteqn{i\hbar \frac{\partial}{\partial t} \psi \left(\vec{r}, t \right)= } \\
& & \left[\sum_{i=1}^N\left(-\frac{\hbar^2}{2m^{\ast}} \nabla^2_{r_i} +V(\vec{r}_i)\right)
+ \frac{e^2}{4\pi \epsilon} \sum_{i>j=1}^N \frac{1}{r_{ij}} \right] \psi \left(\vec{r}, t \right)
\end{eqnarray*}
The above equation can be solved efficiently using a Chebychev
polynomial expansion\cite{Wang1999,Green2004}
\begin{eqnarray*}
 \lefteqn{\psi \left(\vec{r}, t \right)=} \\
& &  \exp\left(-i(E_{\max}+E_{\min}) t /2 \right)
  \sum J_n (\alpha) T_n(-i \tilde{H})\psi \left(\vec{r}, 0 \right),
\end{eqnarray*}
where $E_{\max}$ and $E_{\min}$ are the upper and lower bounds on the
energies sampled by the numerical grid, $J_n(\alpha)$ are Bessel
functions of the first kind, $T_n$ are the Chebyshev polynomials. The
normalized Hamiltonian is defined as $\tilde{H} = \left(2H - E_{\max}
- E_{\min} \right)/(E_{\max} - E_{\min})$ to ensure convergence.

\begin{figure}[htb]
    \includegraphics[width=7cm, bb=0 0 460 680,clip]{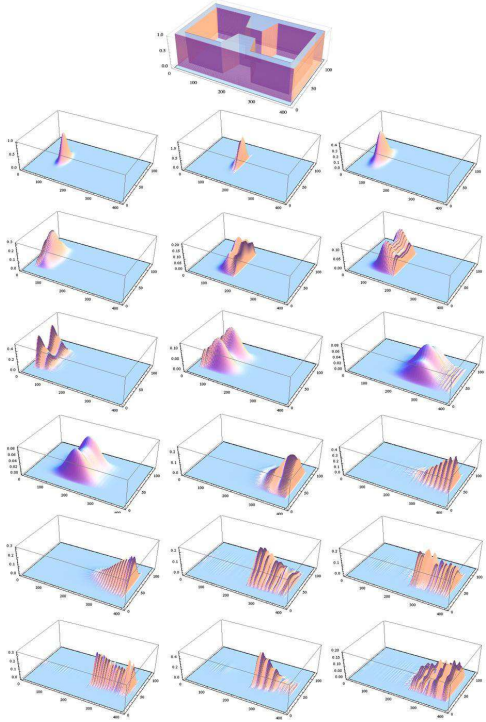}
    \caption{Time evolution of system wave-function in coupled quantum dot system.}
    \label{fig.qubit-simulations}
\end{figure}

\begin{figure}[htb]
    \includegraphics[width=7cm, bb=0 0 440 620,clip]{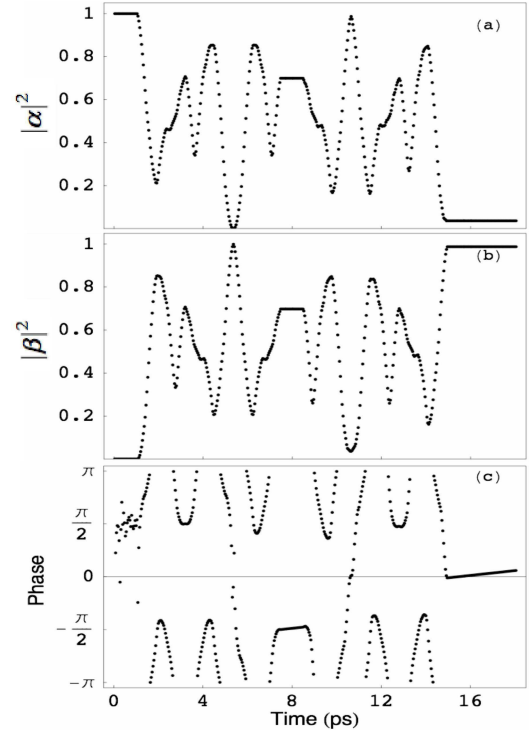}
    \caption{Qubit rotation in the Bloch sphere as time varies.}
    \label{fig.qubit-simulations-2}
\end{figure}

Using such a theoretical framework, all electrons in the system are treated on
equal footing, and they evolve coherently in time under the influence of each other as well as external fields. Fig. \ref{fig.qubit-simulations} illustrates the time evolution of an electron confined in a coupled quantum dot system provided by the above equation. All possible qubit rotations in the Bloch sphere can be accomplished by controlling the central potential barrier and/or electron-electron interactions, as demonstrated in Fig. \ref{fig.qubit-simulations-2} where the amplitudes of and the phase between the  $\ket{L}$ and  $\ket{R}$ states are shown as time varies.

\section{Conclusion}

In this paper, we provided a practical recipe for efficiently implementing quantum random walks on arbitrarily complex graphs using electrons trapped in a 2D array of quantum dots.  The proposed scheme is particularly elegant since the walker is not required to physically step between the vertices of the associated graph, giving rise to a significant advantage over other existing schemes.  Also presented is a detailed simulation of controlled qubit rotation in the Bloch sphere.  Our numerical results show that even the simplest gate operations induce rather intricate quantum dynamics.  Nonetheless, with precision control of the electrodes that define the coupled quantum dots, arbitrary qubit rotations in the Bloch sphere can be accomplished.

\begin{theacknowledgments}
  This work was supported by the Australian Research Council and the University of Western Australia. We also thank R. Green for his earlier work on qubit rotations.
\end{theacknowledgments}

\bibliographystyle{unsrt}
\bibliography{qdot-qrw}

\begin{thebibliography}{10}

\bibitem{Shor1997}
P.W. Shor.
\newblock {\em Proceedings of the 35th Annual Symposium on the Foundations of
  Computer Science}, page 124, 1994.

\bibitem{Grover1997}
L.K. Grover.
\newblock {\em Phys. Rev. Lett.}, 79:325, 1997.

\bibitem{DiVincenzo2005}
David~P. DiVincenzo.
\newblock Double quantum dot as a quantum bit.
\newblock {\em Science}, 309:2173, 2005.

\bibitem{Fedichkin2000}
M.~Yanchenko L.~Fedichkin and K.~Valiev.
\newblock {\em Nanotechnology}, 11:387, 2000.

\bibitem{Tanamoto2000}
T.~Tanamoto.
\newblock {\em Phys. Rev. A}, 61:022305, 2000.

\bibitem{Bertoni2000}
R.~Brunetti C.~Jacoboni A.~Bertoni, P.~Bordone and S.~Reggiani.
\newblock {\em Phys. Rev. Lett.}, 84:5912, 2000.

\bibitem{Hollenberg2004}
C.~Wellard A. R. Hamilton D. J. Reilly G. J.~Milburn L.~Hollenberg, A.
  S.~Dzurak and R.~G. Clark.
\newblock {\em Phys. Rev. B}, 69:113301, 2004.

\bibitem{Green2004}
R.~Green and J.~B. Wang.
\newblock Quantum dynamics of gate operations on charge qubits.
\newblock {\em J. Comput. Theor. Nanosci.}, 1:309, 2004.

\bibitem{Dinneen2008}
C.~Dinneen and J.~B. Wang.
\newblock Quantum dynamics of gate operations on charge qubits.
\newblock {\em J. Comput. Theor. Nanosci.}, 5:1, 2008.

\bibitem{Sessionsa1997}
Richard~B. Sessionsa, Mark Orama, Mark~D. Szczelkuna, and Stephen~E. Halforda.
\newblock Random walk models for {DNA} synapsis by resolvase.
\newblock {\em J. Mol. Bio.}, 270:413, 1997.

\bibitem{Benichou2005}
O.~B\'{e}nichou, M.~Coppey, M.~Moreau, P-H. Suet, and R.~Voituriez.
\newblock Optimal search strategies for hidden targets.
\newblock {\em Phys. Rev. Lett.}, 94:198101, 2005.

\bibitem{Trautt2006}
Zachary~T. Trautt, Moneesh Upmanyu, and Alain Karma.
\newblock Interface mobility from interface random walk.
\newblock {\em Science}, 314:632, 2006.

\bibitem{Kilian2003}
Lutz Kilian and Mark~P. Taylor.
\newblock Why is it so difficult to beat the random walk forecast of exchange
  rates?
\newblock {\em J. Int. Eco.}, 60:85, 2003.

\bibitem{Hoshino1971}
Satoshi Hoshino and Kozo Ichida.
\newblock Solution of partial differential equations by a modified random walk.
\newblock {\em Numer. Math.}, 18:61, 1971.

\bibitem{Ceperley1986}
David Ceperley and Berni Alder.
\newblock Quantum monte carlo.
\newblock {\em Science}, 231:555, 1986.

\bibitem{Berg1993}
Bernd~A. Berg.
\newblock Locating global minima in optimization problems by a random-cost
  approach.
\newblock {\em Nature}, 361:708, 1993.

\bibitem{Scholl2003}
{Joachim Sch\"{o}ll and Elisabeth Sch\"{o}ll-Paschingerb}.
\newblock Classification by restricted random walks.
\newblock {\em Pattern Recognition}, 36:1279, 2003.

\bibitem{Anteneodo2007}
C.~Anteneodo and W.~A.~M. Morgado.
\newblock Critical scaling in standard biased random walks.
\newblock {\em Phys. Rev. Lett.}, 99:180602, 2007.

\bibitem{Bar-Yossef2006}
Z.~Bar-Yossef and M.~Gurevich.
\newblock Random sampling from a search engine's index.
\newblock In {\em WWW '06: proceedings}, pages 367--376. ACM Press, 2006.

\bibitem{Aharonov1993}
Y.~Aharonov, L.~Davidovich, and N.~Zagury.
\newblock Quantum random walks.
\newblock {\em Phys. Rev. A}, 48:1687, 1993.

\bibitem{Knight2003-1}
Peter~L. Knight, Eugenio Rold\'{a}n, and J.~E. Sipe.
\newblock Quantum walk on the line as an interference phenomenon.
\newblock {\em Phys. Rev. A}, 68:020301, 2003.

\bibitem{Kempe2003}
J.~Kempe.
\newblock Quantum random walks: an introductory overview.
\newblock {\em Contemp. Phys.}, 44:307, 2003.

\bibitem{Douglas2008}
Brendan~L Douglas and Jingbo Wang.
\newblock A classical approach to the graph isomorphism problem using quantum
  walks.
\newblock {\em J. Phys. A}, 41:075303, 2008.

\bibitem{Shenvi2003}
Neil Shenvi, Julia Kempe, and K.~Birgitta Whaley.
\newblock Quantum random-walk search algorithm.
\newblock {\em Phys. Rev. A}, 67:052307, 2003.

\bibitem{Childs2004}
A.~Childs and J.~Goldstone.
\newblock Spatial search by quantum walk.
\newblock {\em Phys. Rev. A}, 70:022314, 2004.

\bibitem{Ryan2004}
C.~A. Ryan, M.~Laforest, J.~C. Boileau, and R.~Laflamme.
\newblock Experimental implementation of a discrete-time quantum random walk on
  an nmr quantum-information processor.
\newblock {\em Phys. Rev. A}, 69:012310, 2004.

\bibitem{Di2004}
Tiegang Di, Mark Hillery, and M.~Suhail Zubairy.
\newblock Cavity qed-based quantum walk.
\newblock {\em Phys. Rev. A}, 70:032304, 2004.

\bibitem{Agarwal2005}
G.~S. Agarwal and P.~K. Pathak.
\newblock Quantum random walk of the field in an externally driven cavity.
\newblock {\em Phys. Rev. A}, 65:032310, 2005.

\bibitem{Travaglione2002}
B.~C. Travaglione and G.~J. Milburn.
\newblock Implementing the quantum random walk.
\newblock {\em Phys. Rev. A}, 65:032310, 2002.

\bibitem{Francisco2006}
D.~Francisco, C.~Iemmi, J.~P. Paz, , and S.~Ledesma.
\newblock Simulating a quantum walk with classical optics.
\newblock {\em Phys. Rev. A}, 74:052327, 2006.

\bibitem{Zou2006}
Xubo Zou, Yuli Dong, and Guangcan Guo.
\newblock Optical implementation of one-dimensional quantum random walks using
  orbital angular momentum of a single photon.
\newblock {\em New Jour. Phys.}, 8:81, 2006.

\bibitem{Zhang2007}
Pei Zhang, Xi-Feng Ren, Xu-Bo Zou, Bi-Heng Liu, Yun-Feng Huang, and Guang-Can
  Guo.
\newblock Demonstration of one-dimensional quantum random walks using orbital
  angular momentum of photons.
\newblock {\em Phys. Rev. A}, 57:052310, 2007.

\bibitem{Kovsik2005}
Jozef Ko\v{s}\'{i}k and Vladim\'{i}r Bu\v{z}ek.
\newblock Scattering model for quantum random walks on a hypercube.
\newblock {\em Phys. Rev. A}, 71:012306, 2005.

\bibitem{Joo2006}
Jaewoo Joo, P.~L. Knight, and Jiannis~K. Pachos.
\newblock Single atom quantum walk with {1D} optical superlattices.
\newblock {\em J. Mod. Opt.}, 54, 2007.

\bibitem{Eckert2005}
K.~Eckert, J.~Mompart, G.~Birkl, and M.~Lewenstein.
\newblock One- and two-dimensional quantum walks in arrays of optical traps.
\newblock {\em Phys. Rev. A}, 72:012327, 2005.

\bibitem{Manouchehri2008}
Kia Manouchehri and Jingbo Wang.
\newblock Quantum walks in an array of quantum dots.
\newblock {\em J. Phys. A}, 41:065304, 2008.

\bibitem{Solenov2006}
Dmitry Solenov and Leonid Fedichkin.
\newblock Continuous-time quantum walks on a cycle graph.
\newblock {\em Phys. Rev. A}, 73:012313, 2006.

\bibitem{Manouchehri2008-2}
K.~Manouchehri and J.B. Wang.
\newblock Quantum random walks without walking.
\newblock 2008.
\newblock submitted to PRL.

\bibitem{Kendon2005}
Viv Kendon and Barry~C. Sanders.
\newblock Complementarity and quantum walks.
\newblock {\em Phys Rev A}, 71:022307, 2005.

\bibitem{Edelman2007}
Alan Edelman and Brian~D. Sutton.
\newblock The beta-jacobi matrix model, the cs decomposition, and generalized
  singular value problems.
\newblock {\em Found. Comput. Math.}, 2007.

\bibitem{Wang1999}
J.~B. Wang and S.~Midgley.
\newblock {\em Phys. Rev. B}, 60:13668, 1999.

\end{thebibliography}

\end{document}